# Substrate transfer and *ex situ* characterization of on-surface synthesized graphene nanoribbons

*Gabriela Borin Barin[1], Andrew Fairbrother[1†], Lukas Rotach[1], Maxime Bayle[2†], Matthieu Paillet[2], Liangbo Liang[3,4], Vincent Meunier[4], Roland Hauert[5], Tim Dumslaff[6], Akimitsu Narita[6], Klaus Müllen[6], Hafeesudeen Sahabudeen[7], Reinhard Berger[7], Xinliang Feng[7], Roman Fasel[1,8]* and *Pascal Ruffieux[1]*

[1]Empa, Swiss Federal Laboratories for Materials Science and Technology, nanotech@surfaces Laboratory, 8600 Dübendorf, Switzerland

[2]Laboratoire Charles Coulomb (L2C), Université de Montpellier, CNRS, Montpellier, France

[3]Center for Nanophase Materials Sciences, Oak Ridge National Laboratory, Oak Ridge, Tennessee 37831, USA

[4]Department of Physics, Applied Physics, and Astronomy, Rensselaer Polytechnic Institute, Troy, New York 12180, USA

[5]Empa, Swiss Federal Laboratories for Materials Science and Technology, Joining Technologies and Corrosion Laboratory, Überlandstrasse 129, 8600 Dübendorf, Switzerland

[6]Max Planck Institute for Polymer Research, Ackermannweg 10, D-55128 Mainz, Germany

[7]Chair of Molecular Functional Materials, Department of Chemistry and Food Chemistry, Technische Universität Dresden, Mommsenstrasse 4, Germany

[8]Department of Chemistry and Biochemistry, University of Bern, Freiestrasse 3, CH-3012 Bern, Switzerland

†current address: Engineering Laboratory, National Institute of Standards and Technology (NIST), Gaithersburg, MD 20899, USA

†current address : Institut des Matériaux Jean Rouxel (IMN), Université de Nantes, CNRS, 2 rue de la Houssinière, BP 32229, 44322 Nantes cedex 3, France

**Abstract**

Recent progress in the on-surface synthesis of graphene nanoribbons (GNRs) has given access to atomically precise narrow GNRs with tunable electronic band gaps that makes them excellent candidates for room-temperature switching devices such as field-effect transistors (FET). However, in spite of their exceptional properties, significant challenges remain for GNR processing and characterization. This contribution addresses some of the most important challenges, including GNR fabrication scalability, substrate transfer, long-term stability under ambient conditions and *ex situ* characterization. We focus on 7- and 9-atom wide armchair graphene nanoribbons (i.e, 7-AGNR; and 9-AGNR) grown on 200 nm Au(111)/mica substrates using a high throughput system. Transfer of both, 7- and 9-AGNRs from their Au growth substrate onto various target substrates for additional characterization is accomplished utilizing a polymer-free method that avoids residual contamination. This results in a homogeneous GNR film morphology with very few tears and wrinkles, as examined by atomic force microscopy. Raman spectroscopy indicates no significant degradation of GNR quality upon substrate transfer, and reveals that GNRs have remarkable stability under ambient conditions over a 24-month period. The transferred GNRs are analyzed using multi-wavelength Raman spectroscopy, which provides detailed insight into the wavelength dependence of the width-specific vibrational modes. Finally, we characterize the optical properties of 7- and 9-AGNRs via ultraviolet-visible (UV-Vis) spectroscopy.

**Introduction**

Atomically precise graphene nanoribbons have been extensively studied due to their remarkable electronic and optical properties[1–4]. Due to lateral confinement, GNRs exhibit a sizeable electronic band gap which can be tuned by their width and edge topology[1,5,6]. In order to achieve such control over the band gap, atomic precision in the fabrication of the GNRs is indispensable. A bottom-up approach based on surface-assisted covalent coupling of molecular precursors[1] allows for the synthesis of ultra-narrow GNRs with defined edge topology and uniform width[1,6,7] which results in an electronically homogenous material across the entire growth substrate. The possibility of fabricating uniform, atomically precise GNRs with a tunable band gap makes this material promising for nanoelectronic and optoelectronic applications [3,4,8].

Production is, however, just one aspect of a GNR technology, and in order to evaluate and exploit their properties GNRs must be transferred onto semiconducting or insulating substrates. Different transfer techniques have been developed to transfer graphene and other 2D materials, which could potentially be adapted to graphene nanoribbons[9–12]. The most widely used transfer procedure is the polymer assisted transfer (mainly using poly(-methyl methacrylate) (PMMA))[13], in which graphene or GNRs samples are supported by a thin PMMA layer before the growth substrate is etched away. In general, the polymer-supported transfer is an easy-to-handle process, even though the complete removal of the polymer support remains a challenge. Solvent and thermal treatments have been used for this purpose, but such processes can induce thermal stress, which can damage the final transferred material[14]. Since the purity and integrity of the GNR surface is of extreme importance for studying intrinsic properties and for GNR integration in nanoelectronic devices – such as field effect transistors[4,8,15] – alternative methods need to be investigated.

Here we use a polymer-free process, previously reported by Cai *et al.*[16] and Fairbrother *et al.*[4], which allows the transfer of high quality, structurally intact GNRs. Using this transfer approach, we avoid any polymer residue resulting in a contamination-free transfer with fewer defects (wrinkles and tears) and preserved edge topology. Transferring GNRs to arbitrary substrates opens up the possibility of exploring GNR properties by characterization techniques that cannot be applied to metal-supported GNRs where intrinsic GNR properties are masked by the metallic substrate. Within this framework, it is of critical importance to develop reliable characterization protocols that allow monitoring GNR quality and stability upon transfer and device integration.

In this work, we study the morphological and spectroscopic properties of transferred GNRs with atomic force microscopy (AFM), Raman spectroscopy, and Ultraviolet-Visible spectroscopy (UV-Vis). Using AFM we investigate the overall morphology of transferred GNR films and evaluate our polymer-free transfer procedure. Raman spectroscopy results show that GNRs remain structurally intact upon transfer and have a remarkable stability under ambient conditions, as tracked over a 24-month period. Additionally, multi-wavelength Raman measurements of GNRs reveal a unique non-dispersive behavior in a large range of excitation energies. *Ex situ* characterization furthermore allows for unprecedented insights into the optical properties of GNRs. Using multilayer GNR films realized through repeated transfer sequences onto suitable substrates, we determine the fundamental optical transitions of 7- and 9-AGNRs via UV-Vis spectroscopy.

**On-surface synthesis of 7- and 9-AGNRs**

Samples were produced in a stand-alone ultrahigh-vacuum (UHV) system dubbed "GNR reactor" that has been specifically designed to allow for a fully automated, reproducible and high-throughput fabrication of GNR samples (See figure S1 for GNR reactor details). The fast and reproducible fabrication of GNR/Au(111)/mica samples in the GNR reactor is necessary to produce the large number of samples needed for efficient optimization of transfer procedures and *ex situ* characterization of well-defined GNR samples. As a growth substrate we used commercially available Au(111) thin films supported on mica (4x4 mm$^2$, 200 nm Au; PHASIS, Geneva, Switzerland)

9-AGNRs were synthesized from 3',6'-di-iodine-1,1':2',1''-terphenyl (DITP) as the precursor monomer[5], and 7-AGNR from 10,10′-dibromo-9,9′-bianthryl (DBBA)[1]. First, the Au(111)/mica substrate is cleaned in ultra-high vacuum by two sputtering/annealing cycles: 1 kV Ar$^+$ for 10 minutes followed by annealing at 470 °C for 10 minutes. Next, the monomer is sublimed onto the Au(111) surface from a quartz crucible heated to 70 °C (DITP) or 200 ºC (DBBA), respectively, with the substrate held at room temperature. After 4 minutes of deposition (resulting in approximately 1 monolayer coverage), the substrate is heated (0.5 ºC/s) up to 200 °C with a 10 minute holding time to activate the polymerization reaction, followed by annealing at 400 °C (0.5 ºC/s with a 10 minute holding time) in order to form the GNRs via cyclodehydrogenation of the polyphenylene precursors.

**Transfer Procedure**

In order to explore the GNRs' properties under various conditions, we have transferred the GNRs from the Au(111)/mica growth substrate onto different target substrates. In this work we use a polymer-free method[4,16], as illustrated in figure 1, leading to a clean transfer and a low density of tears and wrinkles in the final GNR film[14,17]. First, the GNR/Au(111)/mica samples are floated on aqueous HCl solution (38%, room temperature, 3.5mL) with the GNR/Au film facing up (step I in figure 1). After ~15 minutes the mica cleaves off from the Au film due to the etching of the interface by HCl (step II in figure 1). After the mica is detached, 3mL of acid is removed and 5 mL of ultrapure water is added. This step is repeated 5 times in order to substantially reduce the HCl concentration.

Prior to transfer of the GNRs, any impurities need to be removed from the target substrate. This is achieved through sonication in acetone and ethanol for 10 minutes followed by rinsing with ultrapure water. In the next step, the target substrate is pushed onto the floating GNR/Au(111) film (step III in figure 1), which results in the adherence of the GNR/Au film on the target substrate. At this stage, the Au film is usually not completely flat on the substrate. To increase adhesion between the Au film and the target substrate we apply a drop of ethanol to the top of the Au film (dried at ambient conditions for 5 minutes) followed by a hot plate annealing step at 100 ºC for 10 min (step IV in figure 1, see also figure S2 for step IV details). Finally, 1-2 drops of potassium iodide based gold etchant (potassium iodine, no dilution), is used to etch away the gold film (~5 minutes). The resulting GNR/substrate sample is cleaned by soaking it in ultrapure water for 5 minutes followed by acetone/ethanol rinsing, and finally dried with nitrogen (step V in figure 1). This polymer-free process results in a large area transfer of GNRs onto arbitrary target substrates while fully suppressing the risk of having polymer residues on the transferred GNRs.

X-ray photoelectron spectroscopy (XPS) measurements were performed to ensure the effectiveness of the cleaning process (with organic solvents and ultrapure water after the Au etching step). As shown in figure S3, 9-AGNRs transferred to $CaF_2$ and $Al_2O_3$ have very small iodine contributions corresponding to ca. 0.3 % and 0.2 % of a monolayer, respectively. GNRs transferred to a $SiO_2$/Si substrate showed no clear iodine signal, indication of iodine to be below the detection limit of ca 0.1 % of a monolayer.

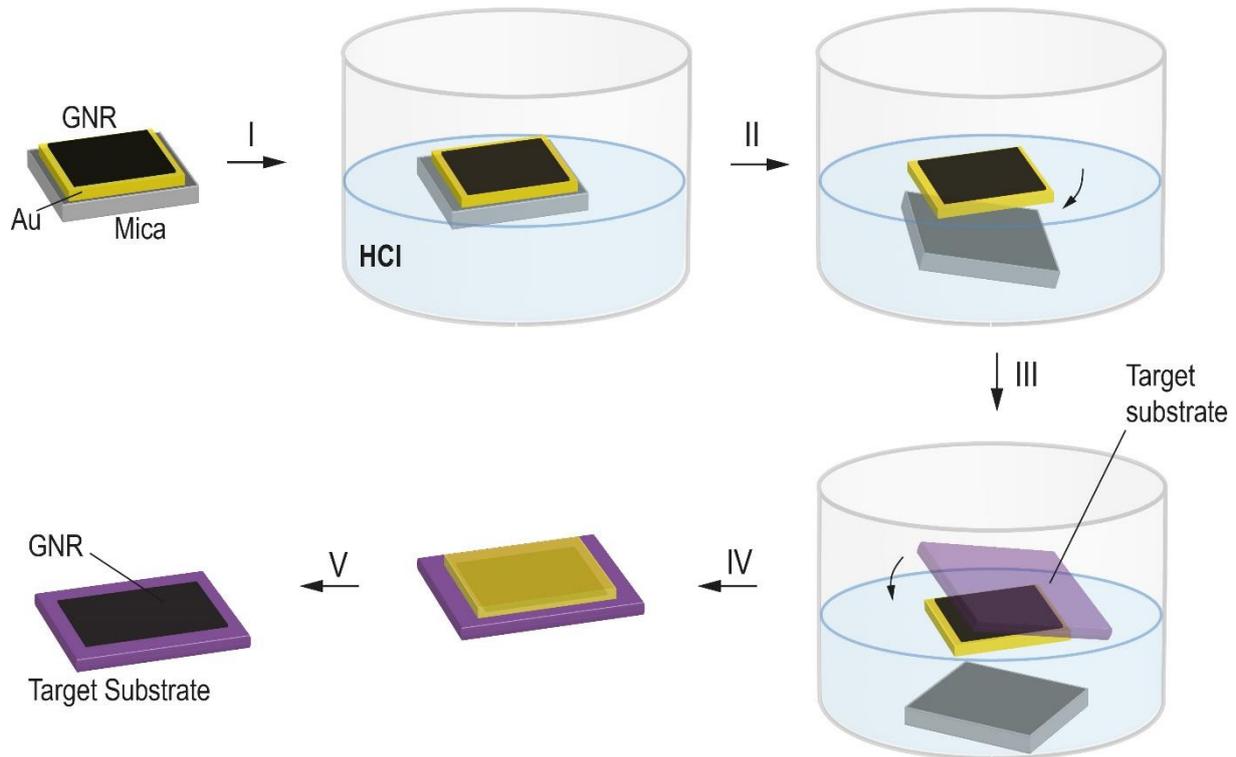

**Figure 1**. Scheme describing the basic steps of the polymer-free GNR transfer procedure: step I: Floating GNR/Au(111)/mica samples on aqueous HCl solution; step II: Cleavage of GNR/Au film from mica substrate (~15minutes); step III: target substrate is pushed onto the floating GNR/Au(111) film; step IV: increase adhesion between the Au film and the target substrate; step V: etching of Au film resulting in GNRs on target substrate.

**Graphene Nanoribbon Characterization**

**Morphology of as-grown and transferred GNR films**

After growth, 9-AGNR samples were removed from the GNR-reactor and inserted in another UHV system for STM characterization. In order to desorb contaminants accumulated during ambient exposure the samples were annealed at 200 °C for 20 minutes. High-resolution STM images taken at room temperature reveal 9-AGNRs with an average length of ~45nm (figure 2(a)). In order to make a direct comparison of the topography of the GNR film before and after substrate transfer, a large scale STM image from a 9-AGNR/Au/Mica sample is displayed in figure 2(b). It reveals the growth uniformity as well as the quasi-continuous film-like morphology that is related to the high coverage preparation.

After transfer to $Al_2O_3$ substrates, atomic force microscopy shows uniform and smooth 9-AGNR-films (roughness ~1-2 nm, see figure 2 (c-d) and figure S4 for roughness profile). High resolution AFM (at ambient conditions) reveals a high coverage of GNRs bundles (2-4 ribbons) across the whole scan area, figure 2(c). As observed in figure 2(d) the transferred GNR film contains relatively few tears, wrinkles, and folds in comparison with polymer assisted transfer[9,14,18]. The observed wrinkles and folds in the GNR film are related to folding of the Au film during the transfer process (see figure S2 for transfer details). This direct comparison of STM and AFM images is the first indication that the polymer-free transfer procedure allows for the uniform transfer of GNRs from Au(111)/mica onto e.g. $Al_2O_3$, which is crucial before exploring other non-vacuum characterization methods and the development of further processing steps for device integration[3].

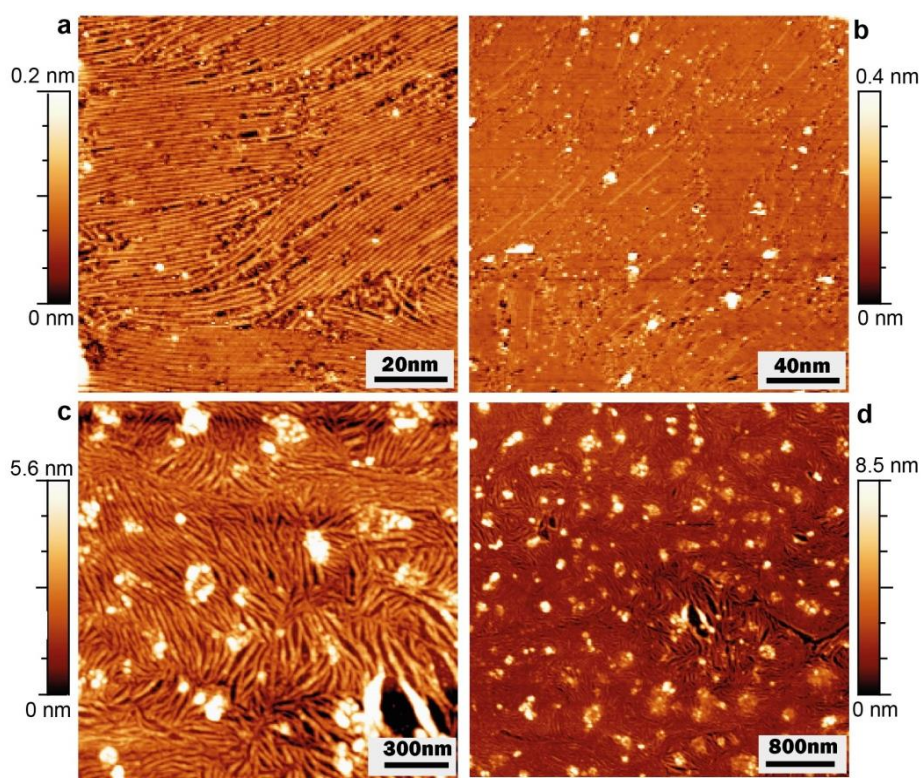

**Figure 2.** (a-b) STM topography images of 9-AGNRs on Au/mica (RT, 1.5V, 0.02nA) and (c-d) AFM topography images of 9-AGNRs transferred onto an $Al_2O_3$ substrate

**Post-transfer stability**

Transferring the nanoribbons to SiO$_2$/Si or other insulating substrates opens up the possibility of investigating the intrinsic properties of unperturbed GNRs. Among the substrates investigated in this work are SiO$_2$/Si substrates that are technologically relevant for device fabrication, Al$_2$O$_3$ and CaF$_2$ that are transparent and hence ideally suited to assess the optical properties of GNRs and transmission electron microscopy (TEM) grids which could potentially be used for atomic scale characterization of transferred GNRs. Raman spectroscopy is a straightforward way to demonstrate the integrity and stability of GNRs after the transfer procedure. The first-order Raman spectra of graphene and other sp$^2$ bonded carbon-based materials present a main band at ~1590 cm$^{-1}$, the so-called G mode[19], which originates from in-plane vibrations and dominates the optical region of graphene-based materials[20–22]. When defects are present, the Raman spectrum of graphene also shows a band at ~1340 cm$^{-1}$, known as the D mode, which is activated by an intervalley double resonance process. Unlike graphene, for which the D band is excited by the presence of defects[23,24], GNRs have a D mode originating from a first-order process[19,25,26]. Due to the edges, the D mode is an intrinsic excitation for GNRs and is expected to be non-dispersive as the G mode[27]. The D mode of GNRs is characterized by a sharp peak (FWHM ~14 cm$^{-1}$) around ~1335-1340 cm$^{-1}$. In addition to a well-structured D mode, GNR spectra also show modes at ~1200-1300 cm$^{-1}$ assigned to the C-H bending modes localized at the edges[28]. The low-energy frequency range also brings valuable information for sp$^2$ carbon with finite width. The radial-breathing mode (RBM) has been widely explored in carbon nanotubes to identify the tube's diameter[29]. A similar mode called radial breathing-like mode (RBLM) was predicted to be related to edge structure and width for graphene nanoribbons[1,25,30]. Sanders *et al.*[31] indicated via tight-binding calculations that the RBLM of GNRs is coherently excited for photon energies near the GNR's lowest optical transitions[5,31]. For the 9-AGNR Prezzi *et al.*[32] calculated this value to be 1.0 eV. For our Raman study, we used a 785 nm laser (1.58 eV), which is sufficiently close to the optical bandgap of 9-AGNRs to yield excitation of the RBLM[5]. In agreement with the prediction by Vandescuren *et al*[30], we observe the RBLM of the 9-AGNR at 311 cm$^{-1}$ [5] on the Au substrate (figure 3(a)). After substrate transfer the RBLM does not show significant changes, indicating that GNRs remain structurally intact upon transfer (figure 3(a)). Moreover, no significant changes are observed for the main peaks of the first order Raman spectra; C-H mode (1232 cm$^{-1}$), D mode (1335 cm$^{-1}$) and G mode (1598 cm$^{-1}$), showing that the GNR's atomically precise edges and the sp$^2$ network are preserved upon substrate transfer.

In addition to demonstrating the robustness of GNRs during transfer, it is important to probe their stability over time under ambient conditions. Achieving long-term stability of GNRs would be a significant step toward their application as an active material in devices. Samples of 9-AGNRs on Au(111)/mica and transferred to SiO$_2$/Si substrates were stored under ambient conditions, and Raman spectra were recorded over a 24-month period. Figure 3(b) shows Raman profiles (average of three spectra) acquired immediately after growth and transfer (0 months, in black), and after 24 months of storage (in red). Compared with the spectra recorded immediately after synthesis on Au(111)/mica and after transfer to SiO$_2$/Si, over time a small change in peak width is observed. For both the as-grown 9-AGNRs on Au(111)/mica and the 9-AGNRs transferred to SiO$_2$, the full-width-at-half-maximum (FWHM) of the D mode is 11 cm$^{-1}$ and increases slightly to 14 cm$^{-1}$ after two years. Similarly, for the G mode we see an increase of the FWHM from 13 cm$^{-1}$ to 16 cm$^{-1}$ which could be associated to adsorbates (atmospheric contamination)[4]. The RBLM mode on Au did not show any significant difference over the 24-month period. For the transferred sample, a detailed analysis of the RBLM mode is difficult since it falls into the region of the Si mode (302 cm$^{-1}$).

For 9-AGNRs, D and G peak shifts remain below the experimental accuracy (3 cm$^{-1}$) of the Raman experiments. Based on these findings, we can affirm that the 9-AGNRs have remarkable stability with no significant structural modifications even after prolonged exposure to air.

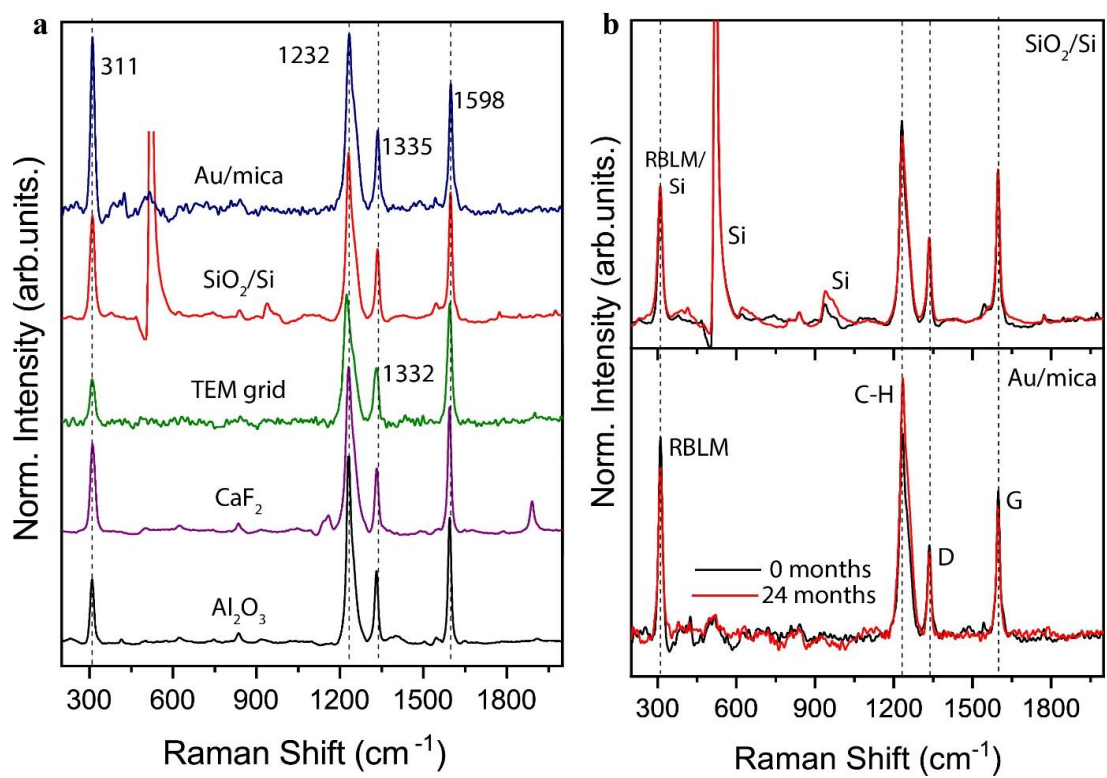

**Figure 3.** Raman analysis of 9-AGNRs on various substrates ($\lambda = 785$ nm). (a) 9-AGNRs on the Au/mica growth substrate and after transfer to SiO$_2$/Si, transmission electron microscopy (TEM) grid, CaF$_2$ and Al$_2$O$_3$ substrates. (b): 9-AGNRs on the Au/mica growth substrate (bottom) and after transfer to a SiO$_2$/Si wafer (top), taken immediately after growth/processing ("0 months", black curves) and after storage under ambient conditions for 24 months (red curves). Average of 3 spectra normalized to the G mode intensity.

**Multiwavelength Raman spectroscopy**

The unique nature of graphene nanoribbons – with their quasi-infinite length, finite width and atomically precise edges – offers a very interesting testbed to investigate well-defined vibrational properties. GNRs with armchair edges and semiconducting behavior do not have Kohn anomalies or linear electronic dispersion [2,20,32]. In the case of cove-GNRs, for example, the $K$ point folds onto $\Gamma$ [33] resulting in a first-order process for activating the D mode, and therefore, a non-dispersive behavior is expected. This latter differentiates the Raman spectra of GNRs from graphene in which the $D$ mode is activated by an intervalley double-resonance process in the presence of defects and shows a strong dispersion in function of the excitation energy[24,25,34]. In order to investigate the (non-)dispersive behavior of the Raman features of AGNRs, both 7- and 9-AGNRs transferred to SiO$_2$/Si substrates were measured with excitation energies ranging from 785 to 457 nm (1.57 eV - 2.7 eV, figure 4 and 5, respectively). The D mode of both 7- and 9-AGNRs is non-dispersive (maximum D mode shift of 5 cm$^{-1}$), indicating the absence of disorder-induced scattering for these substrate-transferred ribbons (see figures S5 and S6 for deconvoluted spectra and figure S7 for dispersion plots). This further confirms that our polymer-free transfer does not induce defects or edge functionalization. In addition, 7-AGNRs reveal non-dispersive behavior for the RBLM peak, observed at 396 cm$^{-1}$ [1,4], and for both C-H related modes at ~1220 cm$^{-1}$ and ~1262 cm$^{-1}$, showing that both width and edge structure are preserved upon transfer. It was not possible to evaluate the dispersion of the RBLM for the 9-AGNR from our experimental data since this mode is only observed with the 785 nm excitation wavelength, which falls nearest to the lowest optical transition of the 9-AGNR (~1eV) [31,32].

Raman spectra measured for both 7- and 9-AGNR were compared to first-principles Raman simulations that take the dependence on the excitation wavelength into account. For both, 7- and 9-AGNRs, we observe an overall agreement between experimental and simulated spectra (figures 4 and 5). Experimental and simulated Raman spectra show a strong dependence on excitation wavelength with respect to both, overall and relative intensities of the different modes. This behaviour originates from the resonant Raman process in which the excitation wavelength overlaps with (or is very close to) an optical transition. Therefore, increased scattering intensities are expected, making it possible to experimentally access the radial breathing-like mode and edge-related modes of 7-AGNRs and 9-AGNRs even at sub-monolayer coverage.

Additionally, for the 9-AGNR we observe a splitting of the G band with a strong dependence of the relative intensities on the excitation energy, for both experimental and simulated spectra. Theoretically, the two G-band components are due to a splitting of the doubly degenerate G mode of extended graphene induced by the finite size and in-plane anisotropy in 9-AGNRs[5]. According to the normal mode analysis (figure S8), this splitting gives rise to two Raman active modes: the G1 mode at 1599 cm$^{-1}$ with $A_g$ symmetry and a second G2 mode at 1623 cm$^{-1}$ with $B_{1g}$ symmetry. For the 7-AGNR, calculations also indicate two G modes with G1 at 1612 cm$^{-1}$ and G2 at 1619 cm$^{-1}$ (figure S8). The simulated separation of G1 and G2 modes for the 7-AGNR is about 7 cm$^{-1}$ and can be observed as a broadening of the G peak measured from 561-785 nm (figure 4 and S5)

Finally, we note that there are some discrepancies in the overall intensity of the peaks between the experimental and simulated spectra. Raman intensities are sensitive to many factors, such as excitonic and substrate effects. In addition, the interface between GNRs and SiO$_2$/Si substrate causes multiple reflections and interference effects[35–37]. These effects are difficult to model and not included in the present simulations, resulting in differences in the relative peak intensities between experimental and simulated spectra. Nonetheless, the simulated spectra provide insights into the intrinsic Raman properties of isolated AGNRs free from external effects.

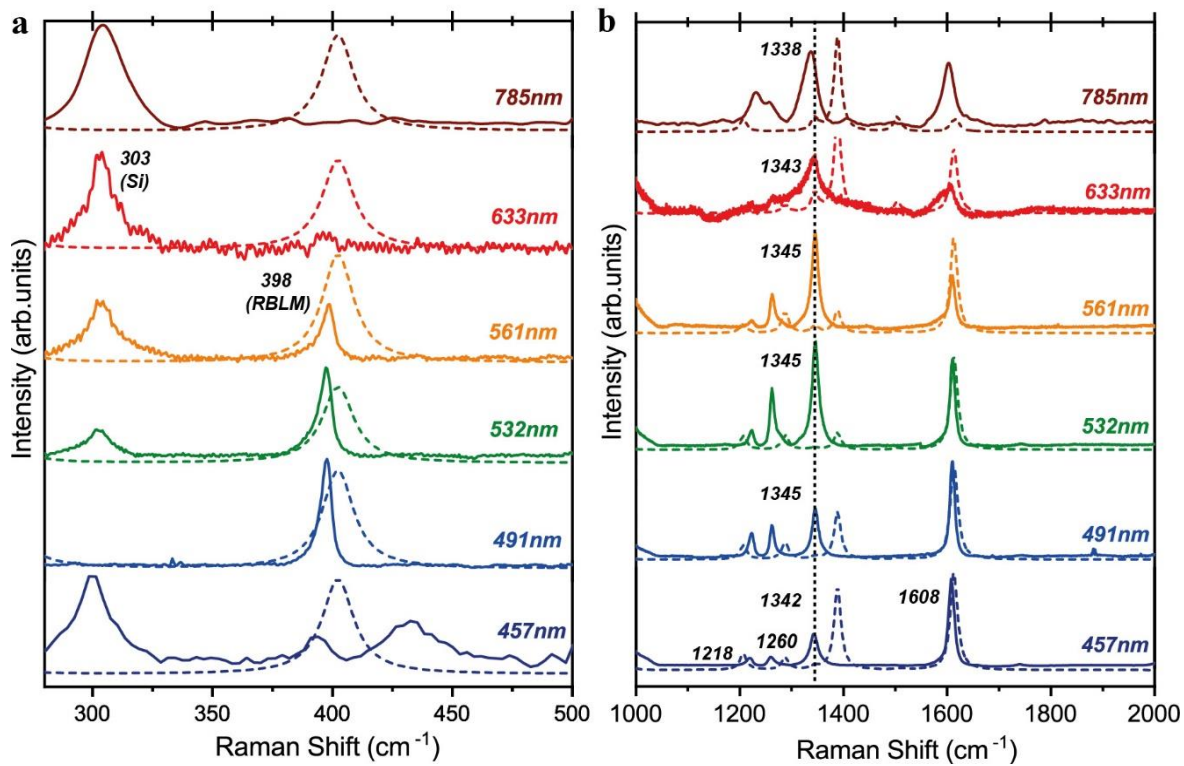

**Figure 4.** Multi-wavelength Raman study of 7-AGNRs (at 0.5 monolayer coverage) transferred to SiO$_2$/Si. (a): low-frequency region. (b): high frequency region (solid line: experimental data; dashed line: simulated spectra). Spectra are not normalized.

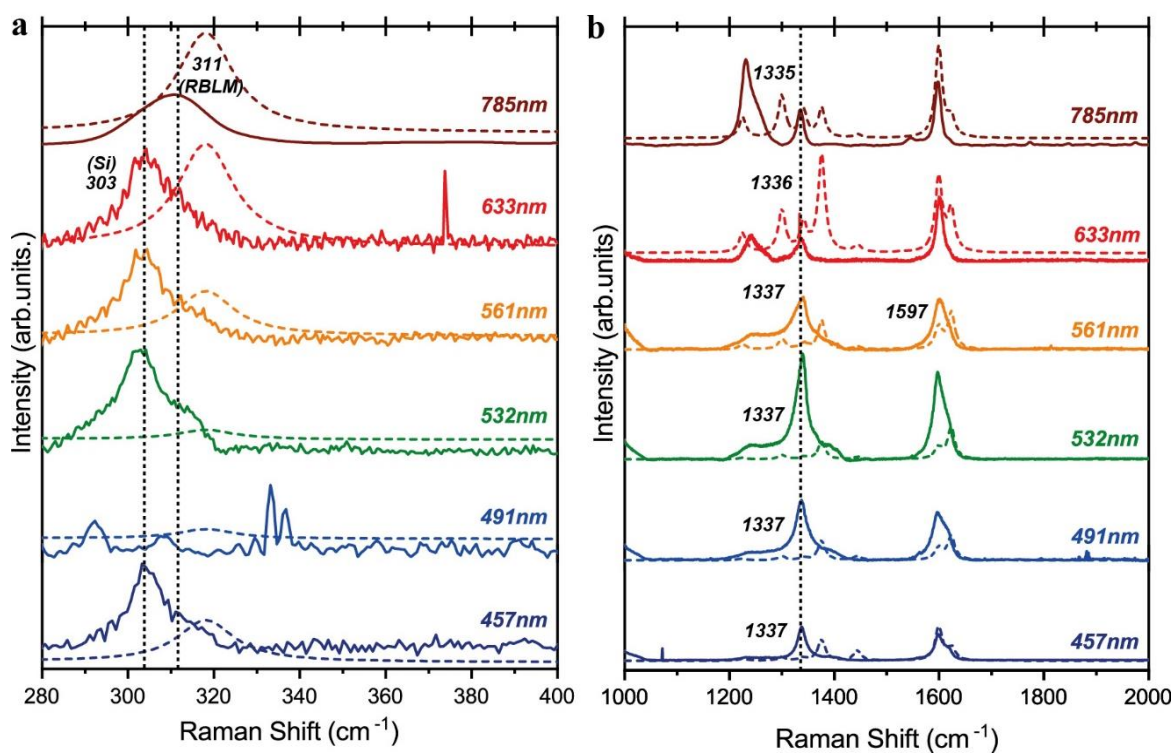

**Figure 5**. Multi-wavelength Raman study of 9-AGNRs (at 0.5 monolayer coverage) transferred to SiO$_2$/Si). (a): low-frequency region. (b) high frequency region (solid line: experimental data; dashed line: simulated spectra). Spectra are not normalized.

**UV-Vis spectroscopy**

Owing to the pronounced width-dependence of the electronic structure of GNRs, the optical absorption can also be tuned over a wide range. For small band gap GNRs, optical absorption can even reach the near-infrared (NIR) region, opening up possibilities for GNRs in technologies such as photovoltaics[33]. Here, 7- and 9-AGNRs have been transferred to (transparent) Al$_2$O$_3$ substrates in order to investigate their absorption spectrum in the UV-Vis range (200 nm to 800 nm) in transmission geometry.

UV-Vis absorption spectra of 7-AGNRs and 9-AGNRs are shown in figure 6. To increase the intensity of absorption features, an Al$_2$O$_3$ sample with 4 layers of GNRs was measured. The four-layer coverage was achieved by a sequential repetition of the transfer process described in figure 1.

For the 7-AGNR we observed strong absorption bands with local maxima at 2.2, 2.4, 4.3 and 5.4 eV (560, 510, 285 and 230 nm respectively). Denk *et al.*[2] computed the optical response of isolated 7-AGNRs which revealed three main absorption features located at 1.9, 2.3, and 4.1 eV. These are in good agreement with the experimental values obtained in this work (figure 6). The two low-energy excitonic contributions at 2.2 eV and 2.4 eV are assigned to optical transitions between the last valence band and first conduction band ($E_{11}$) and between the next to last valence band and second conduction band ($E_{22}$), respectively[2]. In addition, the absorption bands are in very close agreement to the ones determined for metal-adsorbed 7-AGNRs ( 2.1, 2.3 and 4.2 eV)[2], indicating negligible substrate contributions to the optical response of AGNRs.

For the 9-AGNR we observe strong absorption bands between 200 and 600 nm, with local maxima at 212, 233, 270, 347, 440 and 520 nm (i.e., 5.8, 5.3, 4.6, 3.6, 2.8 and 2.4 eV, respectively). It is clear that the first optical excitation predicted to be around 1.0 eV[32] is not accessible for this detection range and that the observed absorption features are thus related to higher order optical transitions. However, the absorption features observed at 2.4 eV, 2.8 eV, and 3.6 eV agree well with the predicted higher-order transitions at 2.4, 2.9, and 3.3 eV[32].

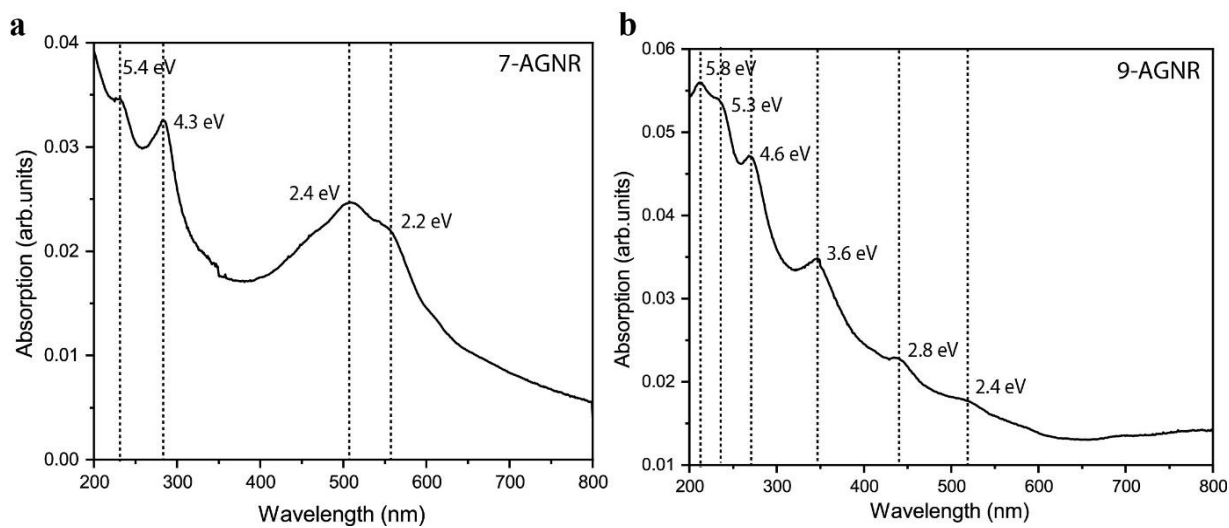

**Figure 6**. (a): UV-Vis spectrum of 4 layers of 7-AGNR transferred to $Al_2O_3$. (b): UV-Vis spectrum of 4 layers of 9-AGNR transferred to $Al_2O_3$.

**Conclusions**

In this work, we report substrate transfer and post-transfer characterization of 7- and 9-AGNRs. AFM imaging reveals that the applied polymer-free transfer procedure yields clean and uniform large area GNR films. Raman characterization before and after GNR substrate transfer indicates that width and edge topology are preserved upon transfer onto different target substrates. In addition, transferred GNRs show remarkable stability over time under ambient conditions, validating their potential for integration in electronic devices. Multiwavelength Raman studies reveal essentially non-dispersive D, G, and RBLM modes in agreement with predictions for defect-free GNRs. Finally, UV-Vis spectroscopy was used to determine the absorption spectrum. For 9-AGNRs, we find strong absorption bands across a wide energy range (6.2 eV – 2.3 eV), while 7-AGNRs present high intensity absorption features at 2.2, 2.4, and 4.3 eV, revealing the marked width dependence of GNR's optical properties. *Ex situ* characterization allowed us to have unprecedented insights on transfer uniformity, stability and quality of transferred GNRs, which are key factors towards the technological application of graphene nanoribbons.

## Materials and Methods

### Au/mica substrates

Au/mica substrates were purchased from Phasis, Geneva Switzerland. They consist of Au (111) films with 200 nm thickness and 4x4 mm$^2$ in dimension, epitaxially grown on muscovite mica sheets by magnetron sputtering. The freshly cleaved mica used as substrate for the Au films has a thickness of around 0.080-0.130 mm. Mica is a silicate material with nearly perfect basal cleavage and is an excellent substrate for growing high-quality quasi-single crystalline Au(111) films.

### Al$_2$O$_3$ and CaF$_2$ substrates

Al$_2$O$_3$ and CaF$_2$ were purchased from Crystal GmbH, Germany. Both substrates were polished on both sides and have dimensions of 7 x 7 x 0.5 mm$^3$. Both substrates were scratch free under magnification of 50x (surface quality given by the provider).

### SiO$_2$/Si substrates

SiO$_2$/Si substrates were purchased from Nova Electronics. They consist of 285 nm dry chlorinated thermal oxide on 525 μm <100> directional Silicon. The 4" wafer was cut in smaller substrates with 7 x 7 mm dimensions.

### TEM grids

TEM grids were purchased from TED PELLA, INC. They consist of monlayer graphene TEM Support Films suspended on a lacey carbon film on a 300 mesh grid.

### Au etchant

Potassium iodide based gold etchant was purchased from Sigma-Aldrich CAS No. 7681-11-0 and used with no further dilution.

### X-ray photoelectron spectroscopy (XPS)

X-ray photoelectron spectroscopy (XPS) was performed in a Physical Electronics Instruments Quantum 2000 spectrometer using monochromatic Al Kα radiation generated from an electron beam operated at 15 kV and 32.3 kV. Spectra were collected under ultrahigh vacuum conditions (residual pressure= $5\times10^{-8}$ Pa) at an analyzer pass energy of 46.95 eV to yield a total analyzer energy resolution of 0.95 eV (for Ag 3d electrons). For compensation of possible sample charging an electron as well as an Ar-ion neutralizer were simultaneously used. The spectrometer energy scale was calibrated for the Au $4f_{7/2}$ signal to be at $84.0 \pm 0.1$ eV

**Scanning tunneling microscopy (STM)**

Topographic scanning tunneling microscopy images of as-grown 9-AGNRs on Au/mica samples were taken with a Scienta Omicron VT-STM operated at room temperature. Constant-current STM images were recorded with 1.5 V sample bias and 0.02 nA setpoint current.

**Atomic force microscopy (AFM)**

The topography of substrate transferred graphene nanoribbon samples was characterized using tapping mode AFM (Bioscope, Bruker) with silicon probes from OPUS (model 160AC-SG Ultrasharp Cantilever ) with tip radius <1 nm (force constant $\sim26$ N m$^{-1}$, resonance frequency in the range of 300 kHz). Height diagrams were recorded with scan sizes of 500 nm, 1.5 and 5 μm and scan speeds of 1 Hz ( 512 × 512 points). The WSxM[38] software was used for AFM analysis.

**Raman Spectroscopy**

Raman spectroscopy with laser wavelength of 785 nm was performed with a Brucker SENTERRA Raman Microscope, operated at 10 mW laser power. Spectra were acquired by 3 sweeps with 10 seconds of integration time and a 50x objective lens.

Multiwavelength Raman spectroscopy (457—633 nm) was performed with an Acton SP2500 spectrometer fitted with a Pylon CCD detector and a grating of 1800 grooves/mm. The samples were excited through a 100x objective (Numerical Aperture 0.9) with power ranging from 0.5-

2 mW and 1 sweep with 10 seconds integration time. The lasers used are at 532 nm a Millennia Prime, Newport, at 457 nm, 491 nm and 561 nm Cobolt DPSS and at 633 nm a He-Ne, Newport.

**Computational procedures**

Plane-wave DFT calculations were performed using the VASP package[39] with local density approximation (LDA) exchange-correlation functionals and projector augmented-wave (PAW) pseudopotentials. The energy cutoff was set at 500 eV. For one-dimensional 9-AGNR and 7-AGNR, the lattice constant in the periodic direction (i.e., $x$ direction) is a = 4.256 Å. In the non-periodic directions ($y$ and $z$ directions), a vacuum region of about 18 Å was used to avoid spurious interactions with periodic replicas. For the unit cell, 24×1×1 k-point sampling was used and all atoms were relaxed until the residual forces were below 0.001 eV/Å. Phonon calculations were then performed using the finite difference scheme implemented in the PHONON software.[40] The Hellmann-Feynman forces in the 3×1×1 supercell were computed using VASP for both positive and negative atomic displacements (δ = 0.03 Å), and then used in PHONON to construct the dynamic matrix, whose diagonalization provides the phonon frequencies and eigenvectors (i.e., vibrations). Raman intensity calculations were then performed within the Placzek approximation[5,41,42] (more details in SI).

**UV-Vis spectroscopy**

UV-Vis spectra were measured on a Cary 5000 UV-Vis instrument with a sample holder with 1 mm mask at room temperature. Absorption/transmission spectra from 200 – 800 nm were recorded at a scan rate of 30 nm min$^{-1}$ (data integral: 1 nm). The spectral bandwidth was set to 5 nm. The substrate reference spectrum was ran separately and later subtracted from the GNRs absorption spectra.

**Acknowledgments**

This work was supported by the Swiss National Science Foundation under Grant No 20PC21_155644, the European Union's Horizon 2020 research and innovation programme under grant agreement number 785219 (Graphene Flagship Core 2), and the Office of Naval Research BRC Program under the grant N00014-12-1-1009. Part of this work (Raman scattering modeling) used resources at the Center for Nanophase Materials Sciences, which is a DOE Office of Science User Facility operated by the Oak Ridge National Laboratory. L. L. was supported by a Eugene P. Wigner Fellowship and by the Center for Nanophase Materials Sciences. Part of the computations was performed using resources of the Center for Computational Innovation at Rensselaer Polytechnic Institute. H.S., R.B., and X.F. were supported by the DFG's Center for Advancing Electronics Dresden (cfaed), EnhanceNano (No. 391979941) and the European Social Fund and the Federal State of Saxony (ESF-Project "GRAPHD"), TU Dresden. T.D., A.N., and K.M. acknowledge the support by the Max Planck Society. G.B.B acknowledges CNPq, Brazil and Empa-Horizon 2020 Marie Skłodowska-Curie action COFUND for financial support. Finally, we acknowledge access to the Scanning Probe Microscopy User lab at Empa for the (AFM) measurements.

# Substrate transfer and *ex situ* characterization of on-surface synthesized graphene nanoribbons

Gabriela Borin Barin, Andrew Fairbrother, Lukas Rotach, Maxime Bayle, Matthieu Paillet, Liangbo Liang,Vincent Meunier, Roland Hauert, Tim Dumslaff, Akimitsu Narita, Klaus Müllen, Hafeesudeen Sahabudeen, Reinhard Berger, Xinliang Feng, Roman Fasel and Pascal Ruffieux

**The GNR reactor**

A dedicated vacuum system (dubbed "GNR reactor") has been set up in order to reduce production time of graphene nanoribbons (GNRs). The GNR reactor allows for a fast and reproducible fabrication of high quality GNR/Au/mica samples, which is necessary to produce enough samples for detailed studies of transfer and device processing. The fully software controlled GNR growth allows fabrication of up to 12 samples per day with minimal operator time expenditure, compared to two or three samples per day on a scanning tunneling microscopy ultra-high vacuum system (STM UHV) with a full time operator.

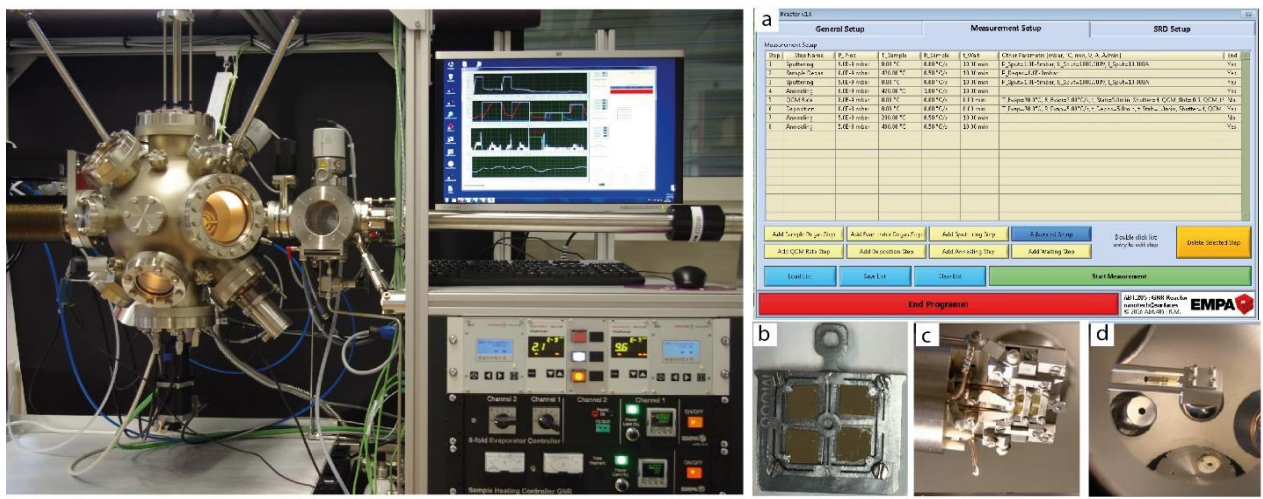

**Figure S1.** Left panel: overview of GNR reactor and electronic rack. Right panel: a) software interface for the definition of the growth steps, b) mounting of up to 4 Au/mica substrates on the sample holder, c) sample transferred onto the sample stage, d) view of quartz crystal microbalance (QCM), shutter of six-fold evaporator and sputter gun.

In a first step, the precursor molecule is introduced into the evaporator, and the growth recipe is set up in the software (LabView) by using existing process element building blocks. The ribbons are grown on 200 nm Au(111) films on mica (Phasis, Switzerland). Up to four 4 mm x 4 mm Au/mica substrates can be mounted on a sample holder, Fig. S1(b). Samples are manually introduced into the fast entry lock chamber, which is then evacuated to a pressure of $\leq 5 \times 10^{-7}$ mbar within 30 minutes using a diaphragm pump and a molecular turbo pump. The substrates are then transferred onto the sample stage in the main chamber with a base pressure of $2 \times 10^{-10}$ mbar, Fig. S1(c). Afterwards, the process recipe can be selected from the GNR reactor software (Fig. S1(a)) and all subsequent mechanical manipulations are fully automated. These steps include sample and quartz microbalance positioning, sample sputtering and annealing steps as well as molecular precursor deposition from a homemade 6-fold evaporator. Detailed experimental parameters used for the growth of 7- and 9-armchair GNRs (AGNRs) are given in the experimental section of the main text.

**Substrate transfer**

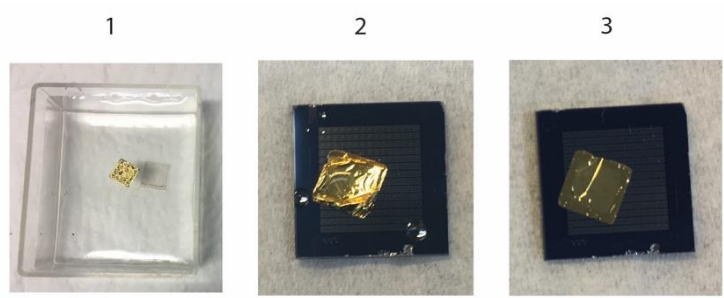

**Figure S2.** Transfer procedure details allowing for decreased surface tension and optimized Au film flatness on the target substrate: (1) After mica cleavage, the Au film floats on the water surface and is ready for pick up by the target substrate. Step (2) shows the Au film right after being picked up by the target substrate (in this case $SiO_2$/Si with Pt pads). At this stage, the film is highly corrugated with a low contact area to the target substrate. (3) In order to increase contact between the Au film and the substrate a two-step process is carried out which includes adding a drop of ethanol, drying under ambient conditions, followed by annealing at 100ºC. Step (3) shows a flat Au film on the target substrate, with a single fold that appears brighter.

# X-ray photoelectron spectroscopy (XPS)

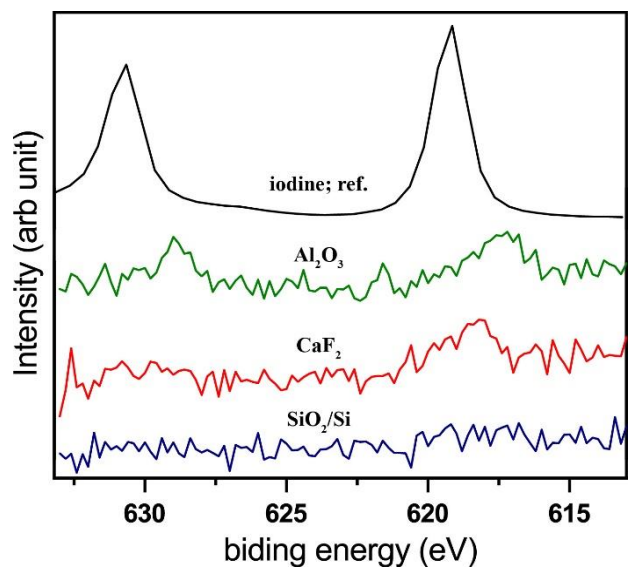

**Figure S3**. XPS measurement of the I 3d core-level region for 9-AGNR samples transferred on SiO$_2$/Si, CaF$_2$, Al$_2$O$_3$ and iodine reference spectrum[1]. 9-AGNR transfer onto these substrates is very clean and results in I 3d core level intensities close to the detection limit of XPS. This indicates that our cleaning step after etching the Au film (with potassium iodine etchant) is efficient and yields essentially iodine-free 9-AGNRs (more details are given in the transfer procedure section of the main text).

# Atomic force microscopy (AFM) – height analysis

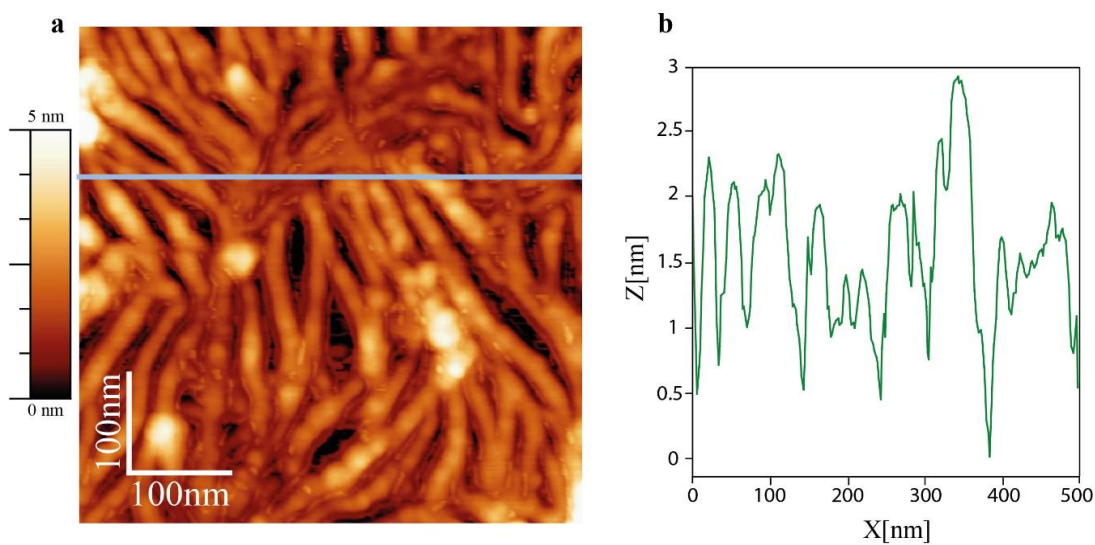

**Figure S4. a)** AFM height image of 9-AGNRs transferred onto $Al_2O_3$ (500 nm scan size), b) height profile extracted from AFM height image (blue line), which shows that 9-AGNR transferred films are smooth and uniform with roughness of the order of 1-2 nm.

**Raman spectroscopy analysis**

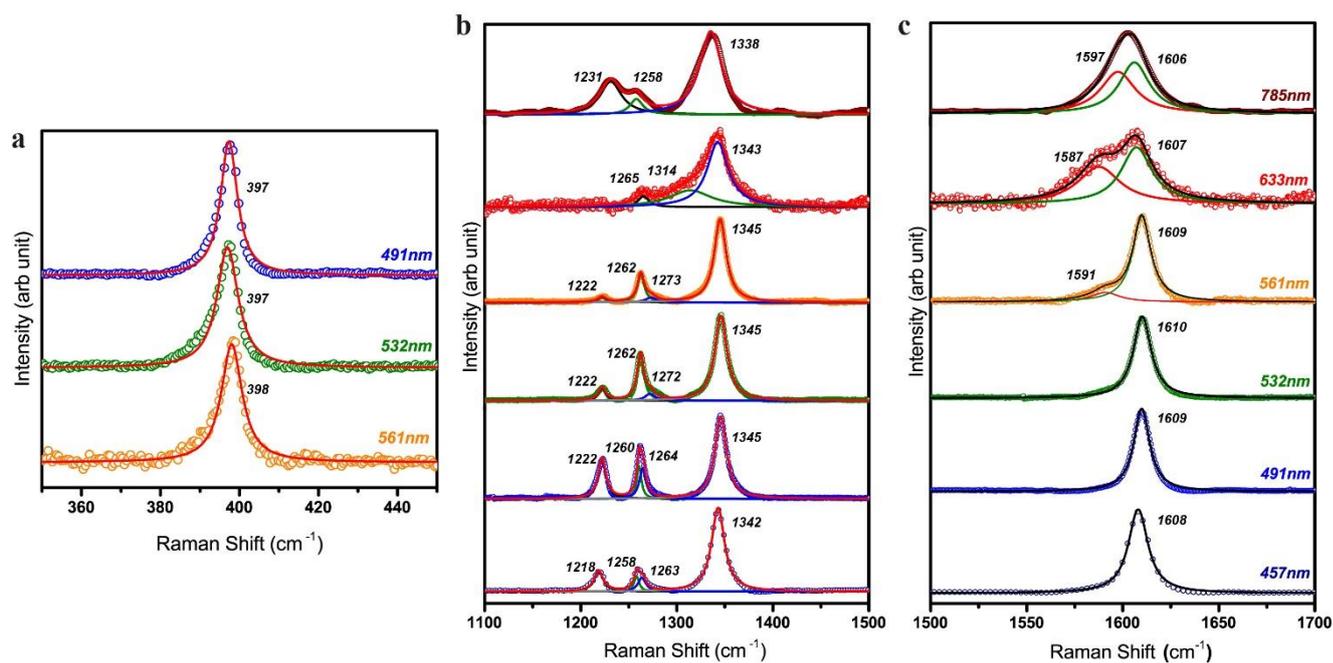

**Figure S5.** Deconvoluted Raman spectra fitted with Lorentzian peaks for 7-AGNR on $SiO_2$/Si substrate: a) radial breathing-like mode (RBLM) b) C-H and D modes, c) G mode.

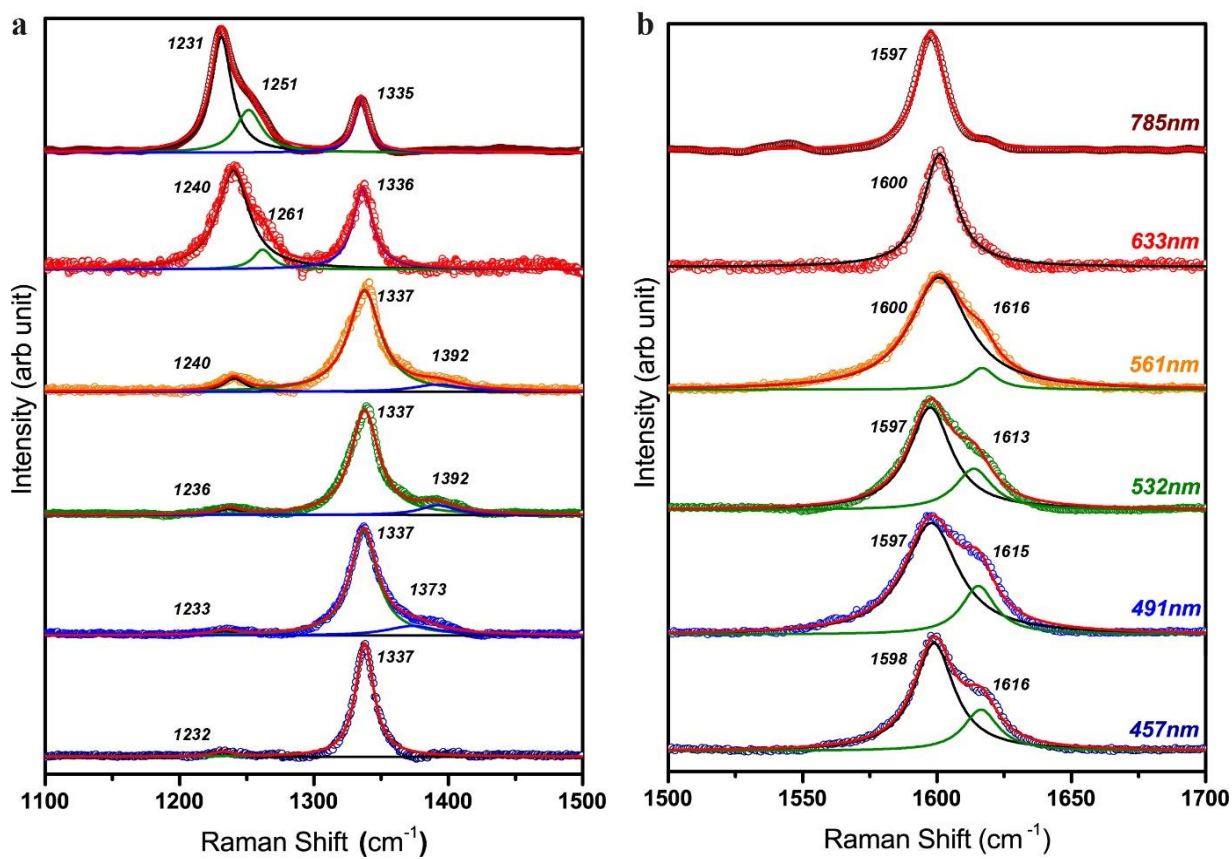

**Figure S6.** Deconvoluted Raman spectra fitted with Lorentzian peaks for 9-AGNR on SiO$_2$/Si substrate a) C-H and D modes, b) G mode.

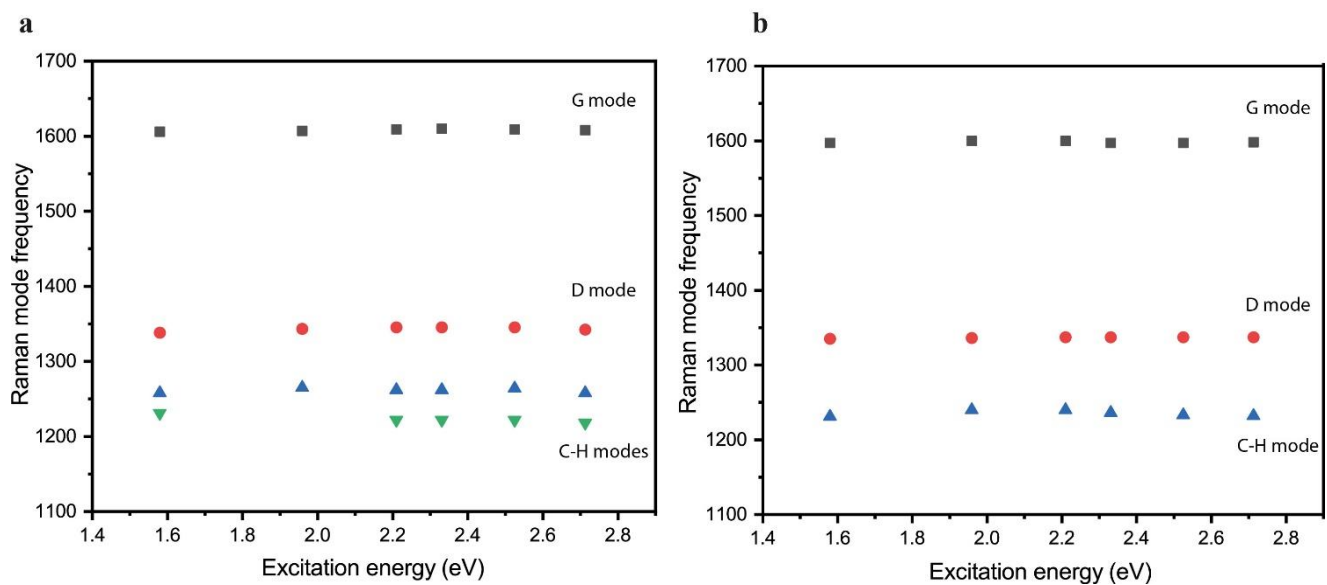

**Figure S7.** Dispersion plots of G, D and C-H modes for 7-AGNR (left panel) excited in the range of 1.5-2.7 eV on SiO$_2$/Si. The G mode is observed at 1608±2, D mode at 1341.5 ± 3.5, CH modes at 1261.5 ± 3.5 and 1226.5 ± 4.5 cm$^{-1}$. Dispersion plots of G, D and C-H modes for

9-AGNR excited in the range of 1.5-2.7 eV on $SiO_2$/Si (right panel). The G1 mode is observed at 1598.5 ±1.5, the D mode at 1336 ± 1 and C-H mode at 1235.5± 4.5 cm$^{-1}$.

**Normal mode analysis**

According to the Placzek approximation, for the $j$-th phonon mode, Raman intensity is $I \propto \frac{(n_j+1)}{\omega_j}|e_i \cdot \tilde{R} \cdot e_s^T|^2$, where $e_i$ and $e_s$ are the electric polarization vectors of the incident and scattered lights respectively, and $\tilde{R}$ is the Raman tensor of the phonon mode. $\omega_j$ is the frequency of the $j$-th phonon mode, and $n_j = (e^{\hbar\omega_j/k_B T} - 1)^{-1}$ is its Boltzmann distribution function at the given temperature $T = 300$ K. The matrix elements of the (3×3) Raman tensor $\tilde{R}$ for phonon mode $j$ at incident laser energy $E_L$ is given by[2-5]

$$\tilde{R}_{\alpha\beta}(j, E_L) = \frac{V_0}{4\pi}\sum_{\mu=1}^{N}\sum_{l=1}^{3}\frac{\partial\varepsilon_{\alpha\beta}(E_L)}{\partial r_l(\mu)}\frac{e_l^j(\mu)}{\sqrt{M_\mu}},$$

where $\varepsilon_{\alpha\beta}(E_L)$ is the complex dielectric tensor at the laser energy $E_L$, $r_l(\mu)$ is the position of atom μ along direction $l$ ($x$, $y$ or $z$), $\frac{\partial\varepsilon_{\alpha\beta}}{\partial r_l(\mu)}$ is the derivative of the dielectric tensor with respect to the atomic displacement, $\frac{e_l^j(\mu)}{\sqrt{M_\mu}}$ is the eigen-displacement of the μ-th atom along the direction $l$ in the $j$-th phonon mode, $e_l^j(\mu)$ corresponds to the eigenvector of the dynamic matrix, $M_\mu$ is the mass of the μ-th atom, and $V_0$ is the unit cell volume. For both positive and negative atomic displacements (δ = 0.03 Å) in the unit cell, the frequency-dependent dielectric tensors $\varepsilon_{\alpha\beta}(E_L)$ were computed by VASP and then their derivatives were obtained via the finite difference scheme. Since the Kohn-Sham gap of conventional density functional theory (DFT) approximations, such as the local-density approximation (LDA), tends to underestimate the optical gaps of GNRs, in the computation of $\varepsilon_{\alpha\beta}(E_L)$ the HSE06 hybrid functional was used. We also note that for non-resonant Raman scattering simulation, it is common practice to neglect the dependence on the incident laser energy and replace the dielectric tensor $\varepsilon_{\alpha\beta}(E_L)$ by the static dielectric constant $\varepsilon_{\alpha\beta}(0)$ at zero laser energy. Here, $\varepsilon_{\alpha\beta}(E_L)$ is computed at the incident laser energy, allowing us to capture Raman resonance effects in first-order Raman scattering[2,4,5]. Finally, based on the phonon frequencies, phonon eigenvectors and the derivatives of dielectric tensors, the Raman tensor $\tilde{R}$ of any phonon mode can be obtained. In the experimental backscattering laser geometry, the polarization vectors of incoming and scattered light are in the $x$-$y$ plane. Averaging over all possible in-plane polarizations, the Raman intensity of any given mode is given by $I \propto \frac{1}{4}\frac{(n_j+1)}{\omega_j}(|\tilde{R}_{11}|^2+|\tilde{R}_{12}|^2+|\tilde{R}_{21}|^2+|\tilde{R}_{22}|^2)$. With the calculated Raman intensities $I(j)$ and phonon frequencies $\omega_j$, the Raman spectrum can be obtained after Lorentzian broadening.

**a** 9-AGNRs

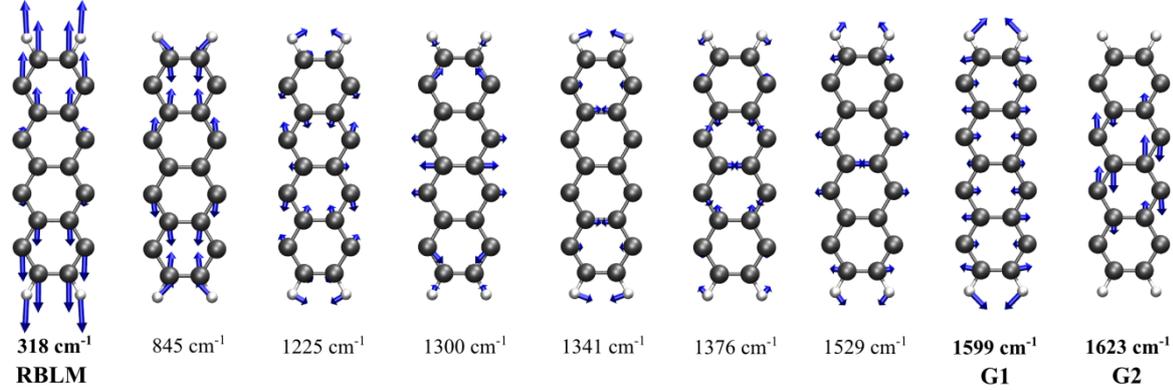

**b** 7-AGNRs

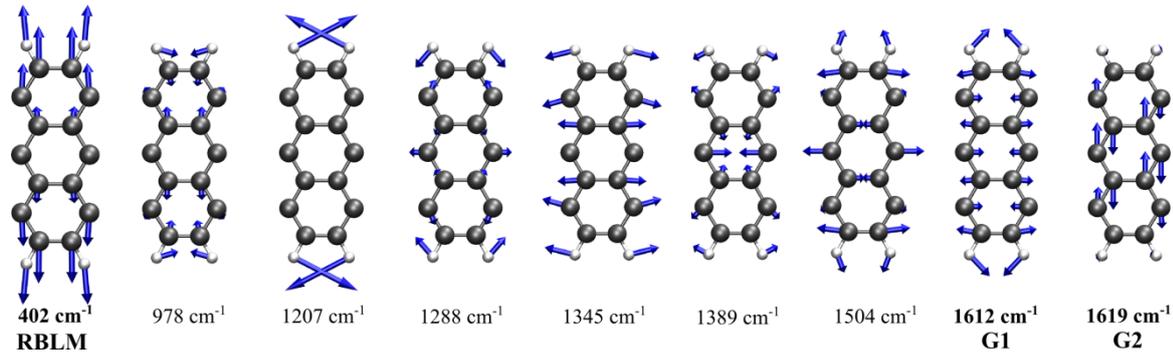

**Figure S8**. Raman normal mode analysis for a) 9-AGNR and b) 7-AGNR (see details on the theoretical procedure). Blue arrows indicate amplitude and direction of atomic displacements. The RBLM, G1 mode and G2 mode are highlighted by bold fonts.